\documentclass[%
 reprint,
 amsmath,amssymb,
 aps,
floatfix,
longbibliography
]{revtex4-2}
\usepackage[dvipsnames]{xcolor}
\usepackage{graphicx}
\usepackage{dcolumn}
\usepackage{bm}
\usepackage{framed}
\usepackage{comment}

\begin{document}

\preprint{APS/123-QED}

\title{Fast generation of time-stationary spin-1 squeezed states by non-adiabatic control}

\author{Lin Xin}
\email{lxin9@gatech.edu}
\homepage{https://physics.gatech.edu/user/lin-xin}
\author{M. S. Chapman}%
\author{T. A. B. Kennedy}
\affiliation{%
  School of Physics, Georgia Institute of Technology, Atlanta, GA 30332-0430, U.S.A\\
}%

\date{\today}

\begin{abstract}

A protocol for the creation of time-stationary squeezed states in a spin-1 Bose condensate is proposed. The method consists of a pair of controlled quenches of an external magnetic field, which allows tuning of the system Hamiltonian in the vicinity of a phase transition. The quantum fluctuations of the system are well described by quantum harmonic oscillator dynamics in the limit of large system size, and the method can be applied to a spin-1 gas prepared in the low or high energy polar states.

\end{abstract}

\maketitle


\section{Introduction}

Creation and characterization of quantum squeezed and entangled states in atomic Bose-Einstein condensates (BECs) with internal spin degrees of freedom are frontier problems in the field of quantum-enhanced measurement and in the investigations of quantum phase transitions and non-equilibrium many-body dynamics \cite{Smerzi2018,MA201189}. 
Condensates with ferromagnetic spin-dependent collisional interactions exhibit a second-order quantum phase transition,  which is tunable by using external fields and available to low-noise tomographic quantum state measurement. 
Experimental studies of collisionally-induced spin squeezing in condensates have mainly utilized time evolution following a magnetic field quench from an initially uncorrelated state to below the quantum critical point (QCP). The squeezing is a result of the quenching and subsequent dynamics generated by the the final Hamiltonian, which is either of the one-axis twisting form \cite{Kitagawa1993,Gross2010} or a close variant \cite{Hamley2012,Muessel2015}. Spin squeezed states have also been generated without quenching through the QCP by parametric/Floquet excitation \cite{Hoang2016,PhysRevLett.125.033401}. 

In addition to these inherently non-equilibrium methods, there is much interest in utilizing adiabatic evolution in spin condensates to create non-trivially entangled ground states such as Dicke states and twin-Fock states \cite{Zhang2013}. Towards this end, there have been experiments using adiabatic \cite{Hoang2017} or quasi-adiabatic \cite{Luo620,Zou6381} evolution across the symmetry breaking phase transition to create these exotic entangled states.  Although some of the interest in these methods has been stimulated by potential applications to adiabatic quantum computing, there are also compelling applications to quantum enhanced metrology \cite{doi:10.1080/0950034021000011536}. A key feature of these approaches is that the entanglement is created in the time-stationary states of the final Hamiltonian, at least in the limit of perfect adiabaticity.

Here, we focus on Gaussian spin squeezed states and consider methods to create time-stationary spin squeezing in a spin-1 condensate by tuning the system Hamiltonian through a pair of quenches of the external magnetic field. Similar squeezed states have previously been discussed in the context of spin-$1/2$ systems \cite{PhysRevA.60.2351,RevModPhys.73.307,Steel1998,Ma2009}. 
Our protocol effectively shortcuts the adiabatic technique \cite{TORRONTEGUI2013117}, overcoming the challenge of maintaining adiabaticity in the neighborhood of the QCP where the frequency scale of the final Hamiltonian evolution tends to zero. Our protocol can be applied to spin-1/2 systems as well, for example,  bosonic Josephson junctions \cite{Laudat2018}. As the squeezing is time-stationary it may be observed directly  without the need for balanced homodyne \cite{PhysRevLett.55.2409} nor Fock states population detection methods \cite{PhysRevLett.76.1796}.

Effectively we propose to implement the quantum harmonic oscillator symplectic Heisenberg picture dynamics  $(X,P) \mapsto M(t) (X,P),$
for times $t$ greater than the squeezed state preparation time $T,$ where
$$
M(t) = \left(
\begin{array}{ccc}
 - x_f \sin \omega_f (t-T)  &  x_f \cos \omega_f (t-T)   \\
- x_f^{-1}\cos \omega_f (t-T)  &  -x_f^{-1} \sin \omega_f (t-T)  \\ 
\end{array}
\right),
$$
satisfies $\det M(t) = 1,$ and $(X,P)$ is regarded as a column vector of initial Heisenberg/Schr\"{o}dinger picture operators \footnote{ Or as classical Hamiltonian phase space coordinates}. Here $\omega_f$ is the oscillator final frequency, i.e., at the end of the protocol, while $x_f$ is the final oscillator dimensionless length scale, defined below. The Heisenberg-limited squeezing of position and momentum variables, associated with the reciprocal factors in the rows of matrix $M(t),$ is independent of time for $t > T$ under this dynamics. Our shortcut protocol requires a preparation time $T\sim\sqrt{\eta}$ which is $\sqrt{\eta}$ faster than the lower limit of the adiabatic passage $T_{ad}\sim \eta$. Here $1/\eta$ is the degree of squeezing of the position or momentum variance. The method solves the problem connecting the time evolution between two quantum oscillator ground states, and its simplicity renders optimal control considerations somewhat superfluous (Appendix \ref{section: optimal control}).  We note that optimal control between canonical thermal states of collections of oscillators has been extensively investigated \cite{B816102J,Andresen_2011,PhysRevE.87.062106}.


In this paper we consider the dynamics of a spin-1 condensate in a magnetic field oriented along the $z$ direction and satisfying the single spatial mode approximation to be described by the Hamiltonian \cite{Hamley2012},
 \begin{equation}
\hat{H}=\frac{c}{2N}\hat{S}^2-\frac{q}{2}\hat{Q}_{z},
\label{hamiltonianreduced}
\end{equation}
where  $\hat{S}^2=\hat{S}^2_{x}+\hat{S}^2_{y}+\hat{S}^2_{z}$ and $\hat{S}_\nu=\sum^N_{i=1} \hat{s}^i_{\nu}$ is a collective spin operator with $\hat{s}_\nu$ the corresponding single particle spin-$\nu$ component, and $N$ is the total number of atoms. The operator $\hat{Q}_{z}=-\hat{N}/3-\hat{Q}_{zz}$ is defined in terms of the nematic (quadrupole)  tensor $\hat{Q}_{\nu\mu}=\sum^N_{i=1} \hat{q}^i_{\nu\mu}$, where $\hat{q}_{\nu\mu}\equiv\hat{s}_{\nu}\hat{s}_{\mu}+\hat{s}_{\mu}\hat{s}_{\nu}-(4/3)\delta_{\nu\mu}$ is a symmetric and traceless rank-2 tensor. The coefficient $c/(2N)$ is the collisional spin interaction energy per particle, with $c<0$ for $^{87}$Rb dictating a preferred ferromagnetic  ordering (in this paper $c = - |c|$ always), while $q=q_z B^2$ is the quadratic Zeeman energy per particle with $q_z / h = 72 $Hz/G$^2$.

We will consider the spin system to be prepared in an eigenstate of $\hat S_z$ with quantum number zero. Since the Hamiltonian commutes with $\hat S_z,$ we may restrict $\hat H$ 
to the subspace of zero net magnetization. The classical phase space corresponds to intersecting unit spheres in the $\{S_{x}, Q_{yz}, Q_z\}$ and $\{S_{y}, Q_{xz}, Q_z\}$ variables. 

\begin{figure}
\begin{center}
\includegraphics{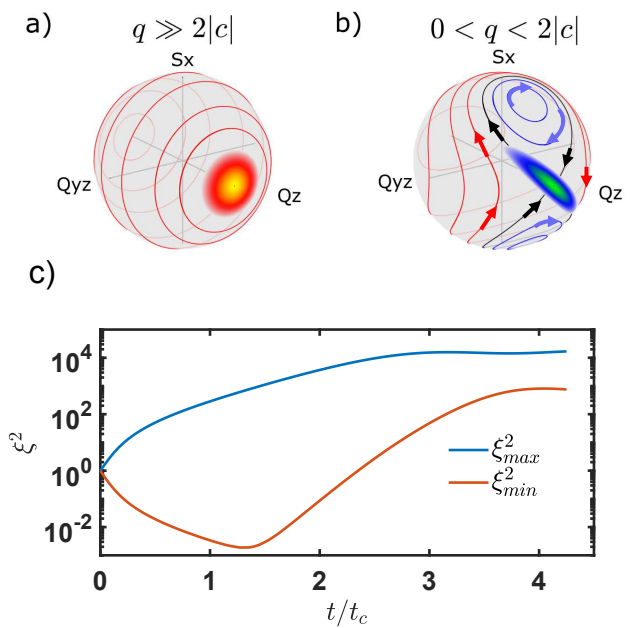}
\end{center}
\caption{(color online). Standard method of creating (non-stationary) squeezed states in a spin-1 BEC. The phase spaces for (a) $q\gg2|c|$, (b) $0<q<2|c|$, respectively. Red/blue lines indicates the energy higher/lower than the separatrix (black lines) across $(\langle \hat{S}_{x}\rangle$,$\langle \hat{Q}_{z}\rangle$,$\langle \hat{Q}_{yz}\rangle$)=(0,N,0) point. The red distribution is an exaggerated illustration of the polar state. The blue distribution is the non-equilibrium evolution for the state initially prepared in (a), which stretches along the seperatrix as observed in our previous work \cite{Hamley2012}. (c) The evolution of the maximum $\xi^2_{max}$ and the minimum squeezing parameter $\xi^2_{min}$ as a function of dimensionless time $t/t_c$. The squeezing is lost after a few characteristic times $t_c=2\pi\hbar/|c|$.}
\label{SN squeezing}
\end{figure}

The classical phase-space orbits of constant energy per particle are shown in Fig. 1. For $q\gg2|c|$, the ground state of the Hamiltonian is the polar state with all atoms having $S_z=0$, and which in the Fock basis can be written as $|N_{+1},N_{0},N_{-1}\rangle=|0,N,0\rangle$, where $N_i$ labels the occupancy of the corresponding Zeeman state, $i=-1,0,1$.  
The polar state gives a symmetric phase space distribution in $\{S_x,Q_{yz}\}$ and $\{S_y,Q_{xz}\}$ as shown in Fig. \ref{SN squeezing}(a) \cite{Hoang2016}. 
 This state is the starting point for many experiments in part because it is easily initialized and stationary in the high $q$ limit. In previous experimental work \cite{Hamley2012}, we have been able to generate a large degree of squeezing by suddenly quenching the magnetic field from $q= \infty$ into the interval $q \in (0,2|c|),$ such that the initial state evolves along the separatix developed in the phase space shown in Fig. \ref{SN squeezing}(b). The time evolution stretches the noise distribution along the separatrix and leads to a large degree of squeezing $\xi_{min}^2 < 1$ for short enough times, as shown in Fig. \ref{SN squeezing}(c). 
 
\begin{figure}
\begin{center}
\includegraphics{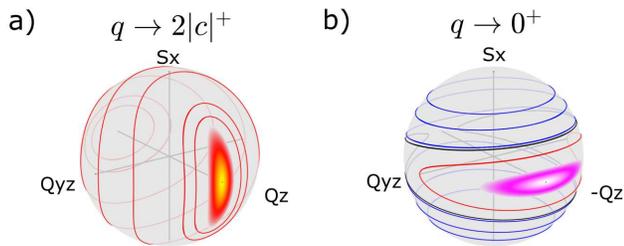}
\end{center}
\caption{(color online). 
Squeezing in the low and high energy polar states. The phase spaces are shown for the low-energy polar state in (a) as $q\to2|c|^+$ (a one-sided limit from above) and in the high-energy polar state in (b) as $q\to 0^+$. As $q$ approaches the QCP, the states follow the deformation of the constant-energy level sets and squeeze accordingly.}
\label{fig: eigen-state squeezing}
\end{figure}

 In this paper, we propose a method to create time-stationary minimum uncertainty squeezed states of a spin-1 condensate. The squeezing is a response to the deformation of the phase space as the system Hamiltonian is tuned close to the QCP. We consider principally the low-energy polar state ($\langle\hat{Q}_z\rangle=1$), whose phase space is shown in Fig. \ref{fig: eigen-state squeezing}(a). We also discuss the experimentally less accessible high-energy polar state ($\langle\hat{Q}_z\rangle=-1$) whose phase space is shown in Fig. \ref{fig: eigen-state squeezing}(b). In both cases, Gaussian fluctuations can be treated by means of quantum harmonic oscillator dynamics in the limit of large particle number.
 
 The remainder of the paper is organized as follows. In section II A we discuss the protocol in the quantum harmonic approximation, proving that time-stationary spin squeezed states are produced by presenting complementary arguments in the Heisenberg and Schr\"{o}dinger pictures. In section II B we consider the validity of the harmonic approximation as a function of system size $N$, by means of numerical solutions of the full dynamics. In section II C we show how the harmonic approximation provides a practical method to estimate the fidelity of state preparation in the experimentally relevant limit of large system size. In Section III we present our conclusions and then several appendices discuss calculation of finite system-size energy gaps useful in estimating residual noise fluctuations, squeezing in the high energy polar state and a brief comparison with the optimal control protocol for our problem. 

\section{Time-stationary squeezing: controlled double quench}\label{section:shortcut}

We consider the condensate to be prepared in the low-energy polar state in the limit of large magnetic field $q \rightarrow \infty.$ (Essential changes in the analysis needed to describe the high-energy polar state are outlined in Appendix \ref{section: highest energy}.) A pair of quenches of the external magnetic field is used to bring the system towards the critical point, where squeezing develops. The procedure thereby avoids the critical slowing down experienced by adiabatic methods \cite{Anquez2016}.

\subsection{Harmonic approximation: $N \rightarrow \infty$} \label{section:harmonic}
In Fig. \ref{fig: eigen-state squeezing}(a) near the pole at $\langle \hat{Q}_z\rangle = 1$, Eq. \ref{hamiltonianreduced} can be approximated by 
\begin{equation}
H= \frac{2c+q}{4}\frac{\hat{S}_x^{2}+\hat{S}_y^2}{N}+\frac{q}{4}\frac{\hat{Q}_{yz}^{2}+\hat{Q}^2_{xz}}{N}+O( \hat{S}_{\mu}^{4},\hat{Q}_{\mu\nu}^4)
\label{harmonic approximation}
\end{equation}
when $q>2|c|$. We see from Eq. \ref{harmonic approximation} that the orbits near the polar axis are harmonic oscillator-like to leading order. The starting point for a quantum harmonic description of the noise fluctuations is the identification of canonically conjugate variables, in the $\hat{S}_z = 0$ subspace formed by $|N_{1},N_{0},N_{-1}\rangle=|k,N-2k,k\rangle =:|k\rangle$, $0\leq k\leq \frac{N}{2}$.

The commutation relations for the subspace are $\langle k'|[\hat{S}_x,\hat{Q}_{yz}]|k\rangle=i(6k-2N)\delta_{k',k}$ and $\langle k'|[\hat{S}_{y},\hat{Q}_{xz}]|k\rangle=i(2N-6k)\delta_{k',k}$.
 Near the $\hat{Q}_z$ pole where $k\ll N$, the commutation relationships are $\langle k|[-\hat{S}_x/\sqrt{2N},\hat{Q}_{yz}/\sqrt{2N}]|k\rangle=i+O(k/N)$ and $\langle k|[\hat{S}_{y}/\sqrt{2N},\hat{Q}_{xz}/\sqrt{2N}]|k\rangle=i+O(k/N)$. Hence
 \begin{align*}
X_1 &:=-\hat{S}_x/\sqrt{2N}, \ \  X_2 :=\hat{S}_{y}/\sqrt{2N}, \\
P_1& :=\hat{Q}_{yz}/\sqrt{2N}, \ \ P_2: =\hat{Q}_{xz}/\sqrt{2N},
   \end{align*}  
    also satisfy  $[X_1,X_2]=[P_1,P_2]=-i\hat{S}_z/(2N)=0, [X_1,P_2]=[X_2,P_1]=i\hat{Q}_{xy}/(2N)=0$ by neglecting terms of $O(k/N)$
   and are thus canonically conjugate variables analogous to a pair of position and momenta. The predictions of the harmonic approximation will be compared with numerical calculations of the full dynamics in section \ref{section: finite N}.
   
   The system is accordingly described by two identical uncoupled quantum oscillators with Hamiltonian
$$
H=\frac{2c+q}{2}(X^2_1+X^2_2)+\frac{q}{2}(P^2_1+P^2_2),
$$
with $[X_{\alpha}, P_{\beta} ] = i \delta_{\alpha,\beta}$. We can identify the effective mass $m = q^{-1}$ and frequency $\omega = \sqrt{q(q-2|c|)}.$ 
As the oscillators are identical and the initial conditions uncorrelated, we treat a generic oscillator in the following discussion and omit the identifying subscript for notational simplicity.

The dimensionless length scale $\sqrt{q / \omega}$ of the oscillator is reciprocal to its momentum scale. The initial condition for the low-energy polar state of the spinor condensate corresponds to the oscillator prepared in its ground state, in which the oscillator position and momentum scales are both equal to unity, $q / |c| >>1.$ In this case the quantum fluctuations are Heisenberg limited and equally shared between position and momentum variables, corresponding to a coherent vacuum state. Regarding the quadratic Zeeman energy $q$ as an external control variable, the target squeezed state is the ground state of a deformed oscillator in which the dimensionless length $x_f = \sqrt{q_f/ \omega_f} $ and momentum $x_f^{-1}$ scales are vastly different. This can be achieved by adjusting $q$ to a final value $q_f$ near to the QCP ($q_c:= 2|c|$), where $\omega_f = 
\sqrt{q_f(q_f - 2 |c|} ) \rightarrow 0 $ and $x_f$ diverges. In the following we propose a procedure which implements the symplectic Heisenberg picture dynamics $(X,P) \mapsto M(t) (X,P),$
for times $t$ greater than the squeezed state preparation time $T,$ where
$$
M(t) = \left(
\begin{array}{ccc}
 - x_f \sin \omega_f (t-T)  &  x_f \cos \omega_f (t-T)   \\
- x_f^{-1}\cos \omega_f (t-T)  &  -x_f^{-1} \sin \omega_f (t-T)  \\ 
\end{array}
\right),
$$
satisfies $\det M(t) = 1,$ and $(X,P)$ is regarded as a column vector of Heisenberg picture operators \footnote{ Or as classical Hamiltonian phase space coordinates}
corresponding to the initially prepared low-energy polar state. As shown in the following subsection the Heisenberg-limited squeezing of position and momentum variables, associated with the reciprocal factors in the rows of matrix $M(t),$ is independent of time for $t > T$ under this dynamics. 

\subsubsection{Heisenberg picture}
The squeezing protocol may be analyzed in the Heisenberg picture as follows. We wish to reduce the Zeeman energy $q$ from large positive values towards $2|c|$, and in order to do this we consider a preparation time $T$ which is bounded at its ends $t=0$ and $t=T$ by a pair of instantaneous quenches in which $q$ is successively reduced through piecewise constant values in the intervals $t < 0$, $0 \le t < T$ and $T \le t.$ We will refer to these regimes as the (low-energy) polar condensate regime, the intermediate regime and the final regime, respectively, and label the oscillator parameters appropriately. The initial value of $q = 10^3|c|$, is used to represent the dominance of the quadratic Zeeman energy in the prepared polar condensate $q/|c| \rightarrow \infty.$ The complete time dependence is given by
$$
q(t) = 10^3|c| \chi_{(-\infty,0) }(t) + q_i \chi_{[0,T) }(t) + q_f \chi_{[T,\infty) }(t),
$$
where the indicator function for a set $A$ is defined by $\chi_A(t) = 1$ for $t \in A$ and $\chi_A(t) = 0$ for $t \notin A,$ and $q_i$ is yet to be determined. We will choose $T$ to be a quarter of the oscillation period of the intermediate oscillator, as in previous discussions of harmonic oscillator squeezing by instantaneous change of frequency \cite{PhysRevA.46.6091, PhysRevLett.67.3665, PhysRevA.79.055804}.

 Just prior to the first quench at $t=0$, the initial Heisenberg picture position and momentum operators are written as $(X,P)$. During the first quench the Heisenberg position and momentum operators are continuous \cite{PhysRevA.20.550,PhysRevA.49.4935}, and 
 with $\omega_i = \sqrt{q_i(q_i-2|c|)},$ the length scale of the intermediate oscillator is $x_i := \sqrt{q_i / \omega_i}.$ For $0 \le t < T$, the time evolution of the Heisenberg operators is then
\begin{align*}
X(t) &= X \cos \omega_i t + x_i^2 P \sin \omega_i t \\
P(t) &= P \cos \omega_i t - x_i^{-2} X \sin \omega_i t .
\end{align*}
 Choosing $T$ to correspond to a quarter period of oscillation $\omega_i T= \pi/2$, gives
\begin{align*}
X(T) =  x_i^2 P  \ \ \mbox{and} \ \
P(T) =  - X /x_i^{2}. 
\end{align*}
This single sudden quench results in a time-periodic oscillation of position/momentum squeezing as observed previously \cite{PhysRevLett.67.3665, doi:10.1080/09500348714550801}.

In the second quench at $t=T,$ $q$ is reduced from $q_i$ to a value $q_f$ closer to $2|c|,$ and the operators $(X(T), P(T))$ are continuous. For $\tau:= t - T \ge 0$,  the subsequent harmonic motion is at the slow frequency $\omega_f = \sqrt{q_f(q_f - 2|c|)},$ and
\begin{align*}
X(t) &= x_f \left[ (x_i^2/x_f) P \cos \omega_f \tau  - (x_f/x_i^2) X \sin \omega_f \tau \right] \\
P(t) &= -x_f^{-1} \left[ (x_f/x_i^2) X \cos \omega_f \tau  + (x_i^2/x_f) P \sin \omega_f \tau \right],
\end{align*}
where $x_f := \sqrt{q_f/\omega_f}$ is the length scale of the final oscillator. 

Now selecting the quadratic Zeeman energies $q_i$ and $q_f$ such that
$$
x_f = x_i^2
$$
gives,
\begin{align*}
X(t) &= x_f \left[  P \cos \omega_f \tau  -  X \sin \omega_f \tau \right] \\
P(t) &= -x_f^{-1} \left[  X \cos \omega_f \tau  + P \sin \omega_f \tau \right],
\end{align*}
corresponding to the desired symplectic transformation $(X,P) \mapsto M(t) (X,P),$ discussed above.
In this case the quantum variances for the initially prepared vacuum state associated with the polar condensate are squeezed, time-independent and Heisenberg limited for $t \ge T,$
$$
\Delta X^2(t) = \frac{1}{2}  \frac{1}{\sqrt{1 - 2 |c| / q_f}} =  \frac{1}{2}  \frac{1}{1 - 2 |c| / q_i}
$$
$$
\Delta P^2(t) = \frac{1}{ 2} \sqrt{1 - 2 |c| / q_f} = \frac{1}{ 2} (1 - 2 |c| / q_i ).
$$
Based on the above analysis, the spin-squeezing parameter in the original variables of the condensate at $q_f\to 2|c|^{+}$ can be described by
\begin{equation}
\xi^{2}_{Q_{yz}}=\frac{\Delta Q^2_{yz}}{N}=\sqrt{1-2|c|/q_f}.
\label{ground state analytic eq}
\end{equation}
We note that to achieve a squeezing variance of $1/ \eta := 1- 2|c| / q_i$ relative to the standard quantum limit, we need to approach within $1/\eta^2$ of the critical point $1-2|c|/q_f = 1/\eta^2$ and $q_i / q_f = 1+ 1/\eta.$

The sensitivity of the squeezing to the condition $x_f = x_i^2,$ between $q_i$ and $q_f$, can be computed by considering a small error $\delta$ in the value of $q_f,$ i.e., $|\delta| / q_f <<1,$
 as the following approximate expression shows,
$$
\Delta P^2(t) \approx \frac{1}{2}  (1 - 2 |c| / q_i) \left[ 1  + \frac{\delta}{q_f} \frac{2 |c| /q_f}{ 1 - 2 |c| / q_f} \sin^2 \omega_f \tau \right].
$$
In the alternative form
\begin{equation}
\Delta P^2(t) \approx  \frac{1}{2 \eta} \left[ 1  + \frac{\delta}{q_f} (\eta^2-1) \sin^2 \omega_f \tau \right],
\label{eqn:sensitivity}
\end{equation}
we observe that to sustain a momentum variance squeezing of $1/(2\eta)$ it is necessary that $$\eta^2 |\delta|/ q_f << 1.$$ A small error in control of the Zeeman energy value in the vicinity of the quantum critical point leads to a quadratic sensitivity to noise fluctuations. 

\begin{figure}
\includegraphics[width=0.5\textwidth]{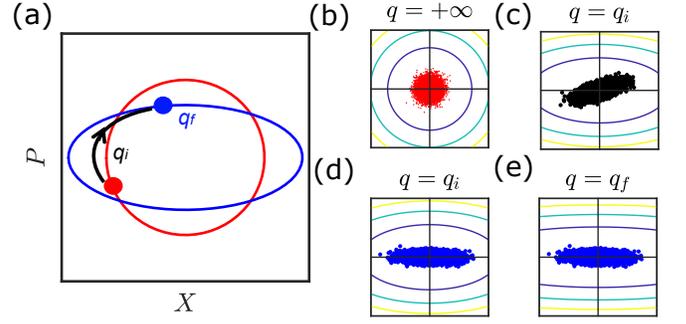}
\caption{(color online). Illustration of the classical phase portrait of the double-quench protocol. (a) $(X,P)$ (red dot) satisfies symmetric condition with an arbitrary phase in the polar state regime. The evolution creates squeezing in the final regime $q_{f}$ orbit (blue solid line) by passing an intermediate $q_{i}$ orbit for a quarter period (black arrow line). (b)-(e) is the ensembles evolution in $(X(t),P(t))$. The initial ensemble (b) satisfies the uncertainty relationships at $q=\infty$.  The ensemble rotates while squeezing in the intermediate regime (c). At $t=T$ (d), a second quench to $q=q_f$ (e) locks the squeezing amplitude and angle. This is in agreement with the quantum harmonic oscillator description for achieving a time-stationary squeezed state. }
\label{cm phase space}
\end{figure}

\begin{figure}
\includegraphics[width=0.5\textwidth]{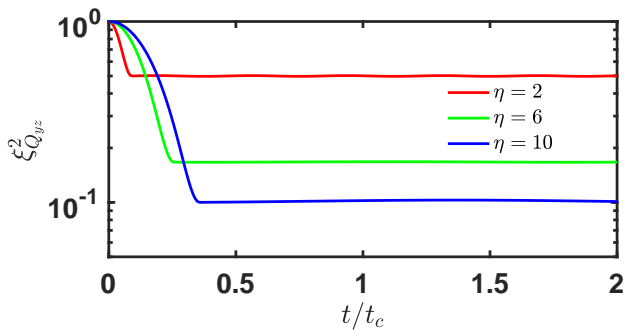}
\caption{(color online). The mean-field numerical simulation result is in good agreement with Eq.  \ref{ground state analytic eq}. A time-stationary squeezing can be generated through a pair of quenches, which maintains the squeezing after a few characteristic times.}
\label{fig: mean-field}
\end{figure}

\begin{figure}
\includegraphics[width=0.5\textwidth]{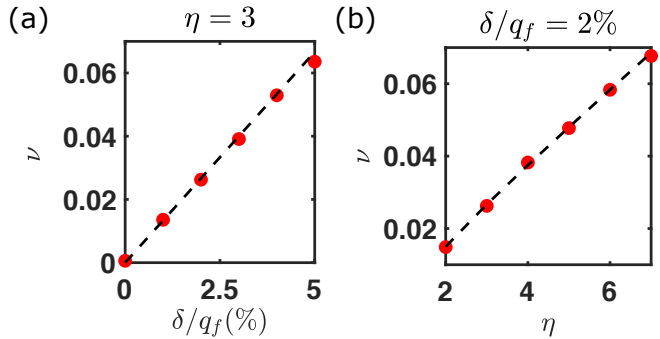}
\caption{(color online). Comparison of Eq. \ref{eqn:sensitivity} with the mean-field simulation dynamics. We plot $\nu=2\max(\Delta P^2(t))-1/\eta$ as the oscillation amplitude of $\sin^2\omega_f\tau$ to illustrate the time-dependent noise fluctuations for $t>T$ when $\delta \ne 0$. (a) $\nu$ as a function of $\delta/q_f$ for $\eta=3$. The red circles are the result from the simulation versus the analytic result (black dashed line). (b) The amplitude is bigger for higher squeezing $\eta$ with $\delta/q_f=2\%$.}
\label{fig:sensitivity}
\end{figure}

We recall that in the $N=+\infty$ limit, the semi-classical dynamics of Eq. \ref{hamiltonianreduced} can be described by the mean-field equations \cite{PhysRevA.72.013602}
\begin{align*}
\dot{\rho}_0&=\frac{2c}{\hbar}\rho_0\sqrt{(1-\rho_0)^2-m^2}\sin(\theta_s)\\
\dot{\theta_s}&=-\frac{2q}{\hbar}+\frac{2c}{\hbar}(1-2\rho_0)+\frac{2c}{\hbar}\frac{(1-\rho_0)(1-2\rho_0)-m^2}{\sqrt{(1-\rho_0)^2-m^2}}\cos(\theta_s),
\end{align*}
where $\rho_0$ is the relative population of $N_0$, $m$ is the relative magnetization and $\theta_s$ is the relative phase between $N_0$ and $N_{\pm1}$. If an ensemble of initial conditions is defined to satisfy the quantum uncertainty relationships $\Delta S_x\Delta Q_{yz}=N$ and $\Delta S_y\Delta Q_{xz}=N$, the numerical simulation (Fig. \ref{fig: mean-field}) shows that the double-quench protocol agrees with Eq. \ref{ground state analytic eq} with a final time-invariant $\xi^2_{Q_{yz}}$. In Fig. \ref{fig:sensitivity},  the noise sensitivity from the numerical simulation is in good agreement with  Eq. \ref{eqn:sensitivity}. 

The classical phase space picture in Fig. \ref{cm phase space}
shows the evolution of an ensemble which passes from the initial state through a quarter period of intermediate dynamics to the final state with the most eccentric level curves.

\subsubsection{Schr\"{o}dinger picture}
In the Schr\"{o}dinger picture the protocol achieves a transformation from the ground state of the initial oscillator, whose quantum fluctuations are those of the polar condensate, to the ground state of the final oscillator created by a pair of quenches. The Schr\"{o}dinger picture operators $X$ and $P$ can be written three separate ways in terms of annihilation and creation operators for the polar condensate $(a,a^{\dagger})$, the intermediate oscillator $(a_i, a_i^{\dagger})$ and the final oscillator $(a_f, a_f^{\dagger})$, using the appropriate oscillator length scales $x_i$ and $x_f$. Hence we can easily see that the oscillator variables are related by an SU(1,1) transformation \cite{PhysRevA.13.2226}
$$
a_i = \mu_i a - \nu_i a^{\dagger}, \ \  \mbox{and}  \ \ a_i^{\dagger} = \mu_i a^{\dagger} - \nu_i a
,$$
where $\mu_i := (x_i + 1/x_i)/2,$ $\nu_i := (x_i - 1/x_i)/2,$ and $\mu_i^2 - \nu_i^2 = 1.$ To achieve the target system discussed above we must control the Zeeman energy quenches so that $x_f = x_i^2,$ and in this special case the final and intermediate oscillator variables are similarly related by 
$$
a_f = \mu_i a_i - \nu_i a_i^{\dagger}, \ \  \mbox{and}  \ \ a_f^{\dagger} = \mu_i a_i^{\dagger} - \nu_i a_i.
$$
Let the vacuum state of the polar condensate oscillator be denoted $| \Phi \rangle$, so that $a | \Phi \rangle = 0$, and the vacuum state of the final oscillator be denoted $| \Omega \rangle,$ so that $a_f | \Omega \rangle = 0.$ The Schr\"{o}dinger picture state vector at time $t=0$ is $|\Psi(0) \rangle = | \Phi \rangle,$ and the double quench produces $|\Psi(t) \rangle = | \Omega \rangle,$ for $t \ge T.$ To see this we introduce the Fock states $\{ | n \rangle \}_{n=0}^{\infty} $ of the intermediate oscillator $a^{\dagger}_i a_i | n \rangle = n |n \rangle.$

Since for each $n=0,1,2,...$
$$
\langle n | \mu_i a_i + \nu_i a_i^{\dagger} | \Phi \rangle = 0
$$
it follows in the sudden approximation, that only even Fock states of the intermediate oscillator are generated in the first quench \cite{agarwal_2012}
$$
\langle 2n | \Phi \rangle = (-1)^n \frac{\sqrt{(2n)!}}{ \sqrt{\mu_i}2^n n!} \left(\frac{\nu_i}{\mu_i}\right) ^n, \ \ n=0,1,2,...
$$
A similar analysis shows that, $\langle 2n | \Omega \rangle=(-1)^n \langle 2n | \Phi \rangle$. Using the sudden approximation also for the second quench, with $\omega_i T = \pi / 2,$ gives 
\begin{align*}
e^{-i \omega_i a^{\dagger}_i a_i T} | \Phi \rangle &= \sum_{n=0}^{\infty} (-1)^n |2n \rangle \langle 2n | \Phi \rangle\\
& = \sum_{n=0}^{\infty} |n \rangle \langle n | \Omega \rangle\\
&= | \Omega \rangle,
\end{align*}
so that the quarter period evolution between quenches prepares the final oscillator ground state, and the time independence of the position and momentum squeezing is readily understood. 

\subsubsection{The high-energy polar state}
In Fig. \ref{fig: eigen-state squeezing}(b) near the pole at $\langle\hat{Q}_z\rangle = -1$, Eq. \ref{hamiltonianreduced} satisfies a quantum harmonic approximation to leading order with mass $m = (q/2)^{-1}$ and frequency $\omega^2 = q(q+2|c|)/4$ in the high-energy polar state case (see Appendix \ref{section: highest energy}).  The same double-quench sequence as described above can also be applied in this case, and the theory goes through with the redefinition of $x_f=\sqrt{q_f/(q_f+2|c|)}$. All of the results  for the low-energy polar state now apply when $|c| \mapsto - |c|$. The squeezing occurs in the position variable rather than the momentum. As a result, the high-energy polar state exhibits spin-squeezing given by
\begin{equation}
\xi^{2}_{S_{x}}=\frac{\Delta S^2_{x}}{N}=\sqrt{q_f/(q_f+2|c|)},
\label{excited state analytic eq}
\end{equation}
as $q_f\to 0^{+}.$

\subsection{Numerical treatment of full squeezing dynamics} \label{section: finite N}
\begin{figure}
\includegraphics[width=0.5\textwidth]{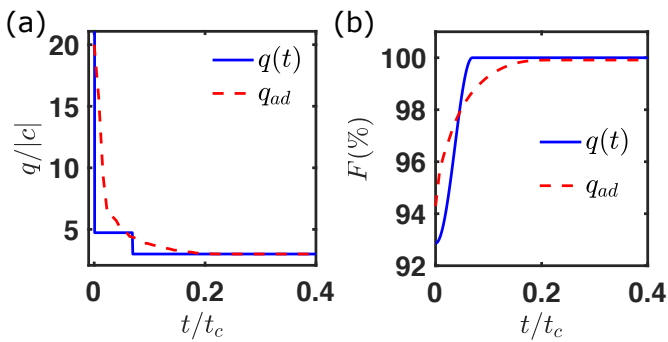}
\caption{(color online). The comparison between the double-quench protocol and the adiabatic passage. (a) $q(t)$ (blue solid line) and $q_{ad}$ (red dashed line) are shown as a function of $t$. (b) Fidelity $F$ is plotted for $q$ (blue solid line) and $q_{ad}$ (red dashed line). Note $q$ reaches higher fidelity of $|\Omega\rangle$ in a shorter time compared to $q_{ad}$.}
\label{cm squeezing comp}
\end{figure}
To assess the limits of validity of the harmonic approximation and the role of finite system size $N$, we numerically solve the full quantum spin-1 dynamics in the single mode approximation
\cite{Law1998},
$$
i\hbar\partial_t|\Psi(t)\rangle=\hat{H}(q(t))|\Psi(t)\rangle$$
 with $\hat{H}(q(t))$ defined by Eq. \ref{hamiltonianreduced}.

We work in a subspace with $\hat{S}_{z}=0$. A suitable basis consists of the states $|N_{1},N_{0},N_{-1}\rangle=|k,N-2k,k\rangle =:|k\rangle$, $0 \leq k \leq \frac{N}{2}$. In this basis, the relevant matrix elements needed to construct the Hamiltonian matrix are
\begin{align*}
\langle k'|\hat{S}_x^2+\hat{S}_y^2|k \rangle &=2\bigg[\Big(2(N-2k)k+(N-k)\Big)\delta_{k',k}+\\
&(k+1)\sqrt{N-2k}\sqrt{N-2k-1}\delta_{k',k+1}+\\
&(k)\sqrt{N-2k+2}\sqrt{N-2k+1}\delta_{k',k-1}\bigg],\\
\langle k'|\hat{Q}_{z}|k \rangle &=4k \delta_{k',k},
\end{align*}
and the initial condition is $|\Psi(t=0)\rangle=|0,N,0\rangle$. The time dependence of the magnetic field quench $q(t)= q_i \chi_{[0,T)}(t) + q_f \chi_{[T,\infty)}(t)$ is used in the simulation with the relation $1+2c/q_f = (1+2c/q_i)^2$ and $\omega_i T = \pi/2$ predicted by the harmonic approximation. The ground state of the final oscillator, denoted $|\Omega\rangle$ above, can be compared to the numerically computed lowest eigenvector of $H(q_f)$.

The quench dynamics may be compared to those of an alternative time-dependence governed by the adiabatic passage function, $t \mapsto q_{ad}(t)$ \cite{Hoang2017}. The latter is determined by taking an initial value of $q=20|c|$ and then optimizing the computed ground state fidelity, defined below, over a set of linear decreasing functions of time in the first time interval. The procedure is then repeated over the sequence of time intervals, to ensure that the fidelity with respect to the instantaneous Hamiltonian ground state at each step exceeds $99.9\%$.
We note that this method outperforms the linear $q$ ramp \cite{Anquez2016}, the Landau-Zener ramp \cite{Zener1932, Zurek2005, Damski2008, Altland2009} and the exponential ramp \cite{Sala2016}.

To verify that we achieve the many-body ground state, the computed fidelity $F=|\langle \Psi(t)|\Omega\rangle|^2$ is shown in Fig. \ref{cm squeezing comp}(b), where $|\Omega\rangle$ is the numerically-computed lowest eigenvector of $H(q_f)$.  In these simulations, the quench function $t \mapsto q(t)$ results in a final fidelity $F$ that satisfies $1 - F < 10^{-6},$ with the squeezing parameter $\eta\approx 2$, $N=10^3$ and $q_f=3|c|$. Although the overall quench sequence is apparently non-adiabatic, we can see the final state is indeed the many-body ground state at $q=q_f$. It is clear that the function $t \mapsto q(t)$ outperforms the optimized adiabatic ramp fidelity and does so in a shorter preparation time. For the squeezing variance factor $\eta$ the preparation time is given by
$$
T=\frac{\pi}{2\omega_i}=\frac{\pi}{2q_f}\frac{\eta}{\eta+1}\sqrt{\eta}.
$$
By comparison, the adiabatic passage preparation time $T_{ad}$ can be estimated by the Landau-Zener  \cite{Zener1932,Zurek2005} and Kibble-Zurek theories \cite{Zurek1985,KIBBLE1980183}, and is bounded by the relaxation time 
$$T_{ad}\geq\frac{2\pi}{\omega_f}=\frac{2\pi}{q_f}\eta.
$$
In Fig. \ref{cm squeezing comp}, this is illustrated for the case of $\eta\approx 2.$

We note that similar numerical computations can be used to investigate the high-energy polar state by employing the initial state $|\Psi(t=0)\rangle=|N/2,0,N/2\rangle$ in the Schr\"{o}dinger equation and the relation $1-2c/q_f = (1-2c/q_i)^2$ in the quench function $t \mapsto q(t)$.

\subsubsection{Numerical investigation of finite $N$ effects}

\begin{figure}
\includegraphics[width=0.5\textwidth]{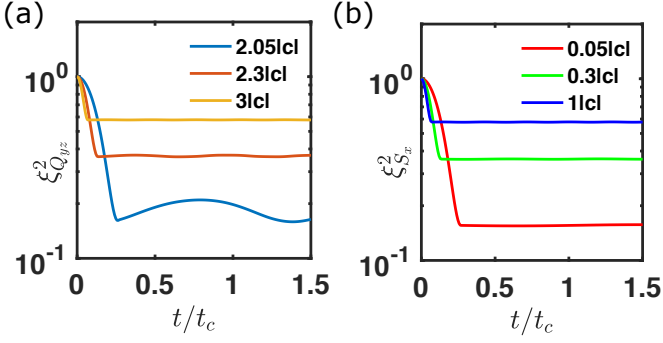}
\caption{(color online). The system-size dependent effect ($N=10^3$) on the shift of the QCP will lead to time-dependent oscillations in the final regime. (a) $\xi^{2}_{Q_{yz}}$ versus $t$ curves are plotted for the low-energy polar state with $q_f\in [2.05|c|,3|c|]$; (b)  $\xi^{2}_{S_{x}}$ versus $t$ curves are plotted for the high-energy polar state with $q_f\in [0.05|c|,1|c|]$. By contrast with (a), there is no time-dependent oscillation as the critical point has no system-size dependent shift.}
\label{cm_overlap}
\end{figure}

\begin{figure}
\includegraphics[width=0.5\textwidth]{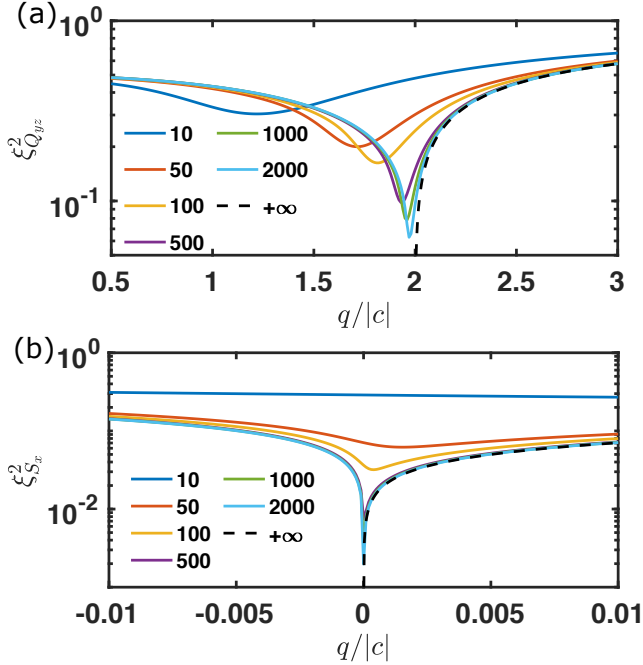}
\caption{(color online).  The finite system-size effect on the maximum squeezing due to the non-zero minimal energy gap. Analytic curves (black dashed lines) are compared with the full quantum simulation of  $N\in[10,2000]$ (solid lines). A sufficiently large $N$ is necessary for the preparation of a highly squeezed state in the low-energy polar state case (a) and the high-energy polar state case (b).}
\label{quad var 10c}
\end{figure}

When we consider finite atom number $N$, the system-size dependent effect becomes important.  For the low-energy polar state the QCP is shifted by an $
N$ dependent quantity
$$2-q_c/|c|\approx e^{3/2}N^{-2/3}$$ 
(see Appendix \ref{eigenstates}). For the high-energy polar state, $q_c=0$ is unshifted. In Fig. \ref{cm_overlap}, we consider $N=10^3$. In Fig. \ref{cm_overlap}(a) for the low-energy polar state, the shift acts as an effective error $\delta$ in the Zeeman energy, as described in section \ref{section:harmonic}, causing time-dependent oscillations in the final regime. In the previous section, we showed the oscillation is suppressed when $\eta^2|\delta|/q_f\ll 1$. $|\delta|\approx 2|c|-q_c=|c|e^{3/2}10^{-2}$ is estimated by the system-size shift. The graphs shown for $q_f=3|c|$ and $q_f=2.3|c|$ both satisfy this condition, while for $q_f=2.05|c|$ we estimate $\eta^2|\delta|/q_f=1.8$. In Fig. \ref{cm_overlap}(b) by contrast, the high-energy polar state shows no oscillation in the final regime due to the absence of a system-size dependent shift.

In the harmonic oscillator description, the maximum squeezing occurs at the QCP where the frequency $\omega_f\to 0$. However, there is a non-zero minimal energy gap $\Delta$ between the ground state and the first excited state \cite{Zhang2013} due to the finite atom number for the low-energy polar state. Thus the maximum squeezing computed by solving the ground state of $H(q_f)$ is limited by $N$ as shown in Fig. 8(a). The same maximum squeezing limit occurs for the high-energy polar state with a non-zero minimal energy gap $\Delta_{E}$ between the highest and the second highest excited states \cite{Qiu2020} (see Appendix \ref{eigenstates}).

\begin{figure}[h!]
\includegraphics[width=0.5\textwidth]{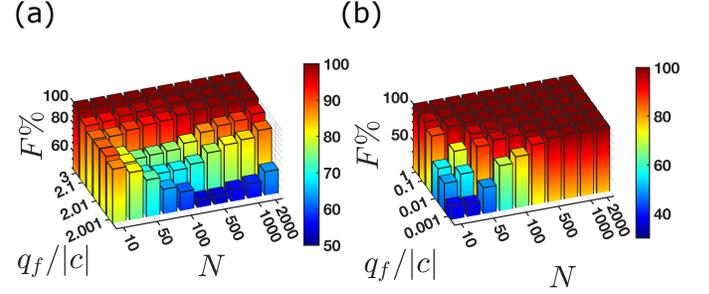}
\caption{(color online). The fidelity of the target state is numerically computed as a function of atom number $N$ and $q_{f}$: (a) low-energy polar state: as $q_{f}$ approaches $2|c|$, large $N$ is necessary but not sufficient for large squeezing due to the shifted QCP. (b) high-energy polar state: as $q_{f}$ approaches $2|c|$, large $N$ is sufficient for large squeezing due to the unshifted QCP.}
\label{cm effect factor}
\end{figure}

The fidelity of target states generated with the double-quench protocol is summarized in Fig. \ref{cm effect factor} as a function of $q_f$ and $N$.

\subsection{Estimating the target state fidelity}
In this section, we discuss a simple practical method to experimentally estimate the fidelity of production of the final oscillator ground state ( $| \Omega \rangle := | 0\rangle$) in the limit of large $N$ (say $N\geq 10^3$). In practice, $N\approx 10^5$ is commonly measured in a spinor condensate of $^{87}$Rb. If the system is prepared in the final target ground state $|0 \rangle$, the squeezing uncertainties will be time-independent. On the other hand, if the system has an admixture of excited states, then the squeezed and antisqueezed quadrature fluctuations will vary harmonically at the frequency $\omega_f$ and produce a modulation in time given by the oscillation of the variances $\mbox{Osc}[\xi^{2}_{\mu}]$, $\mu=S_x,Q_{yz},$ defined below. By a spectral analysis of the numerically-computed wavefunctions, we find that only the ground and the first-excited states are significantly occupied. 

The simplest model which mixes an excited state component to the final oscillator ground state is a coherent superposition in which the first excited state, $|1 \rangle$, has probability $p_e$, $$
|\Psi(t)\rangle=\sqrt{1-p_e}|0\rangle+e^{-i(\omega_f t+\phi)}\sqrt{p_e}|1\rangle,$$ with $\phi$ a relative phase.

We note the matrix elements for the final oscillator at $q=q_f$ are
\begin{equation}
\begin{split}
\langle 0| \Delta X^2 | 0\rangle= -\langle 0| \Delta X^2 | 1\rangle =\frac{1}{2}\sqrt{\frac{q_f}{q_f+2c}}\\
\langle 1| \Delta X^2 | 1\rangle=\frac{3}{2}\sqrt{\frac{q_f}{q_f+2c}}.
\end{split}
\label{eqn:expectation}
\end{equation}
As noted in section \ref{section: finite N},  Eq. \ref{eqn:expectation} may be checked against a full numerical computation of matrix elements of $\hat{S}_x^2/2N=X^2$ and $ \hat{Q}_{yz}^2/2N=P^2$ using the ground and the first excited states of the final Hamiltonian. We have obtained excellent agreement for all matrix elements in the case of $N=10^3$ and $q_f\in(2|c|,10|c|]$.

\begin{figure}
\includegraphics[width=0.5\textwidth]{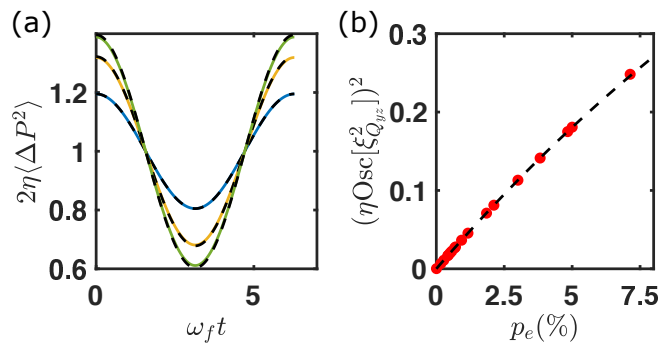}
\caption{(color online). (a) The relative time-dependent oscillation in the final regime is compared with the prediction from the coherent superposition model of Eq. \ref{eqn:Fourier} (dashed lines). $2\eta\langle \Delta P^2\rangle$ are numerical simulated as the solid curves with $\{p_e, q_f\}=\{0.01,2.7\},\{0.03,2.1\},\{0.05,2.05\}$, $N=4000$. (b) The relative oscillation amplitude predicted by the coherent superposition model of Eq. \ref{correlation} (dashed line) and the full quantum simulation show good agreement with each other. The correlation between $(\eta\mbox{Osc}[\xi^{2}_{Q_{yz}}])^{2}$ versus $p_{e}$ with $\eta\in[2,6]$ suggest we can use $\mbox{Osc}[\xi^{2}_{\mu}]$ as an indicator of the target state fidelity.}
\label{delta xi vs ep}
\end{figure}

 The model predicts the mean-square oscillator fluctuations for $|\Psi(t)\rangle$ is given by,
 with as before $1-2|c| / q_f =: 1/\eta^2,$
\begin{equation}
\begin{split}
\langle &\Psi(t)| \Delta X^2 | \Psi(t)\rangle\\
&=\left((1+2p_e)-2\cos(\omega_f t+\phi)\sqrt{p_e(1-p_e)}\right)\frac{\eta}{2}\\
\langle &\Psi(t)| \Delta P^2 | \Psi(t) \rangle\\
&=\left((1+2p_e)+2\cos(\omega_f t+\phi)\sqrt{p_e(1-p_e)}\right)\frac{1}{2\eta}.
\end{split}
\label{eqn:Fourier}
\end{equation}
To compare the simple model with the full numerical
solutions we should first check that the ratio of the mean to the oscillation of the $\cos(\omega t)$ harmonic is independent of $x_f^2 = \sqrt{q_f/(q_f - 2|c|)}$ and depends only on $p_e$. In Fig. \ref{delta xi vs ep}(a), we show that this feature of the model is indeed consistent with full numerical computations.

In Eq. \ref{eqn:Fourier}, the amplitude of $\cos(\omega t)$ is defined as $\mbox{Osc}[\xi^2_\mu]$ and follows the relationship
\begin{equation}
\begin{split}
(\mbox{Osc} [\xi^2_{S_x}]/\eta)^2=(\eta\mbox{Osc} [\xi^2_{Q_{yz}}])^2=4p_e(1-p_e),
\end{split}
\label{correlation}
\end{equation}
where $\xi^2_{S_x}=2\Delta X^2$ and $\xi^2_{Q_{yz}}=2\Delta P^2$. This result may also be compared with numerical computations (Fig. \ref{delta xi vs ep}(b)).  Since in the model $p_e$ is directly related to the fidelity $F$ through $p_e=1-F$, measurement of $\mbox{Osc}[ \xi_\mu^{2}]$ provides an estimate of the fidelity of the final ground state.

\section{Conclusion}
In summary, we have discussed a protocol for the preparation of time-stationary squeezed states in spin-1 BECs. The protocol simply involves a sequence of two reductions in the Zeeman energy of the system in an external magnetic field, in order to tune the system Hamiltonian close to a QCP. The proposed method appears to be simpler and faster than the typical adiabatic techniques.  We also propose a procedure to measure the fidelity of the state preparation by monitoring the harmonic oscillation of the asymptotic squeezing dynamics. 

We expect the methods proposed in this paper may be applied to other similar many-body systems, for example,  anti-ferromagnetic spinor condensates with $c>0$ \cite{Zhao2014, Sala2016}, bosonic Josephson junctions \cite{Laudat2018} and the Lipkin-Meshkov-Glick model \cite{Solinas2008}. We believe that the proposed method is ideal for experiments, and could enable the observation of time-stationary squeezing in spin-$1$ systems for the first time.

\begin{acknowledgments}
We thank M. Barrios and J. Cohen for stimulating insights and discussions. Finally, we acknowledge support from the National Science Foundation, grant no. NSF PHYS-1806315.
\end{acknowledgments}

\appendix
\section{Energy gap for finite system-size  $N$}\label{eigenstates}
\begin{figure}[h]
\includegraphics[width=0.5\textwidth]{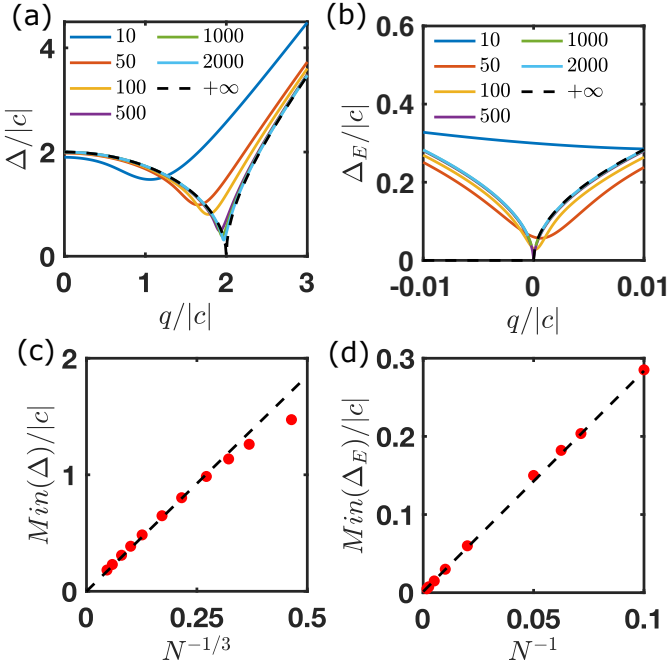}
\caption{(color online). The finite system-size effect on the QCP shift and the minimal energy gap. Energy gap (a) $\Delta$ and (b) $\Delta_{E}$ are simulated for $N\in[10,2000]$. (a) The QCP for the low-energy polar state is system-size dependent. (c) The minimum energy gap $Min(\Delta)$ is proportional to $N^{-1/3}$. (b) The QCP for the high-energy polar state is fixed when the system size varies. (d) The minimum energy gap $Min(\Delta_E)$ is linearly correlated to $N^{-1}$.}
\label{energy gap} 
\end{figure}

\begin{figure}[h]
\includegraphics[width=0.5\textwidth]{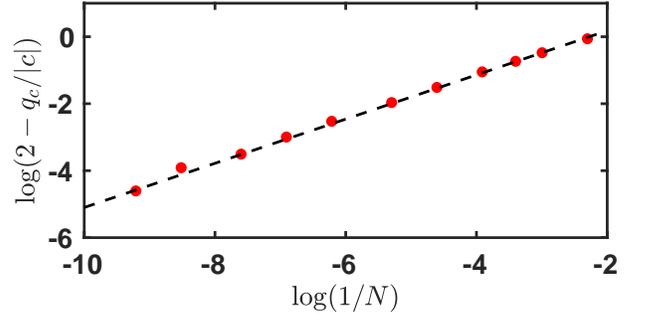}
\caption{(color online). The QCP shift in Fig. \ref{energy gap} (a) is computed as a function of $N$. The dashed curve is $\log(2-q_c/|c|)\approx\log(1/N)/3+3/2$ (the least square linear fit gives $\log(2-q_c/|c|)=0.6467\log(1/N)+1.462$). This gives the estimation of the system-size when applying the double-quench shortcut with good robustness.}
\label{critical point shift} 
\end{figure}

The energy gaps $\Delta$ and $\Delta_{E}$ and lowest energy eigenvectors are computed by numerical diagonalization of $\hat{H}(q)$ as a function of $q$ and $N$ in the Fock state basis as mentioned in section \ref{section: finite N}. The energy gaps are shown in Fig. \ref{energy gap}. Here the low-energy polar state gap is defined to be $\Delta:=E_1-E_0$. Similarly the two highest energy eigenstates can be computed with the energy eigenvalues of $E_{N/2}$ and $E_{N/2-1}$. The high-energy polar state gap is computed as $\Delta_E:=E_{N/2}-E_{N/2-1}$. The value of $q_c$ is system size-dependent for the low-energy polar state case following the approximate relationship $2-q_c/|c|=e^{3/2}N^{-2/3}$ as plotted in Fig. \ref{critical point shift}. From this $q_c$ is computed by finding the location of the minimal energy gap over the simulated range. The quantum phase transition for the low-energy polar states is second order \cite{Xue2018} while for high-energy polar states is first order  \cite{Qiu2020}.

\section{Harmonic approximation for the high energy polar state}\label{section: highest energy}
For $q\gg2|c|$, the initial high-energy polar state of the Hamiltonian is the twin-Fock state, which in the Fock basis can be written as $|N/2, 0, N/2\rangle$.
The twin-Fock state also gives a symmetric phase space distribution in $\{S_x,Q_{yz}\}$ and \{$S_y,Q_{xz}\}$. Near the pole on the $\langle \hat{Q}_z\rangle=-1$, Eq. \ref{hamiltonianreduced} can be approximated by
$$
H= \frac{2c-q}{4}\frac{\hat{S}_x^{2}+\hat{S}_y^2}{N}-\frac{q}{4}\frac{\hat{Q}_{yz}^{2}+\hat{Q}^2_{xz}}{N}+O( \hat{S}_{\mu}^{4},\hat{Q}_{\mu\nu}^4).
$$
The commutation relationships are $\langle k|[-\hat{S}_x/\sqrt{N},\hat{Q}_{yz}/\sqrt{N}]|k\rangle=i+O(N/2-k)/N)$ and $\langle k|[\hat{S}_{y}/\sqrt{2N},\hat{Q}_{xz}/\sqrt{2N}]|k\rangle=i+O((N/2-k)/N)$ when $(N/2-k)\ll N$. The conjugate variables can be hence defined by neglecting the $O((N/2-k)/N)$ terms as
 \begin{align*}
X_1 &:=-i\hat{S}_x/\sqrt{N}, \ \  X_2 :=i\hat{S}_{y}/\sqrt{N}, \\
P_1& :=i\hat{Q}_{yz}/\sqrt{N}, \ \ P_2: =i\hat{Q}_{xz}/\sqrt{N}.
\end{align*}

The quantum fluctuations are again controlled by two identical uncoupled quantum oscillators with Hamiltonian
$$
H=\frac{q-2c}{4}(X^2_1+X^2_2)+\frac{q}{4}(P^2_1+P^2_2).
$$
With $[X_{\alpha}, P_{\beta} ] = i \delta_{\alpha,\beta}$, we can identify the mass $m = (q/2)^{-1}$ and frequency $\omega^2 = q(q+2|c|)/4$. Under this definition, the double-quench treatment can be applied. In this case the quantum variances for an initially prepared twin-Fock state are squeezed, time-independent and Heisenberg limited for $t \ge T,$
\begin{align*}
\Delta X^2(t) &= \frac{1}{2}  \frac{1}{\sqrt{1 +2 |c| / q_f}} =  \frac{1}{2}  \frac{1}{1 +2 |c| / q_i}\\
\Delta P^2(t) &= \frac{1}{ 2} \sqrt{1 + 2 |c| / q_f} = \frac{1}{ 2} (1 + 2 |c| / q_i ).
\end{align*}

\section{Optimal control considerations}\label{section: optimal control}
The optimal control method minimizing the preparation time $T$ through the cost function $J(T) =\int_0^T 1 dt$ proposed for thermal states in \cite{B816102J} also provides the time-optimal solution to the transfer between initial and final oscillator ground states. In our system, the initial polar condensate state is prepared in a large quadratic Zeeman energy $q_0$ before $t=0$. The optimal control sequence is a three step jump between $q_0$ and $q_f$ (or equivalently, between $x_0=\sqrt{q_0/\omega_0}$ and $x_f=\sqrt{q_f/\omega_f}$) characteristic of the so-called ``bang-bang" switching between $\sup q(t)$ and $\inf q(t)$.

\begin{figure}[h]
\includegraphics[width=0.5\textwidth]{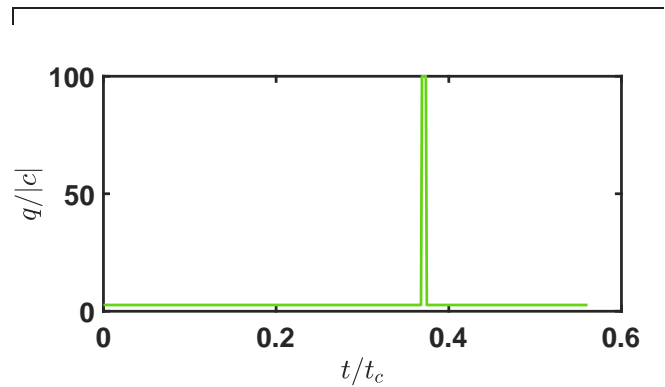}
\caption{(color online). The optimal control function calculated for $q_0=100|c|, q_f=3|c|$. The evolution begins with a switch from $q_0$ to $q_f$ at $t=0$ and then a double switch $q_f \to q_0 \to q_f$ over a very short interval below $t/t_c = 0.4$.}
\label{fig:bang-bang} 
\end{figure}

For the initial condition $x_0=1$ and $\tau:=t-T \ge 0$, 
the phase space map $(X,P) \mapsto M (\tau) (X,P)$ is given by
\begin{widetext}
\begin{align*}
M(\tau)=\frac{x_f}{1+x_f^2}\begin{pmatrix}
x_f\cos(\omega_f\tau)-\sqrt{1+x_f^2+x_f^4}\sin(\omega_f \tau), \sqrt{1+x_f^2+x_f^4}\cos(\omega_f\tau)+x_f\sin(\omega\tau)\\
-x_f^{-1}\sin(\omega_f \tau)-x_f^{-2}\sqrt{1+x_f^2+x_f^4}\cos(\omega_f\tau), -x_f^{-2}\sqrt{1+x_f^2+x_f^4}\sin(\omega_f\tau))+x_f^{-1}\cos(\omega_f \tau)
\end{pmatrix}.
\end{align*}
\end{widetext}
This matrix still satisfies the condition $\Delta X^2(t)=x_f^2/2$ and $\Delta P^2(t)=1/(2x_f^2)$ although unlike our double quench protocol it is not a symplectic transformation. The total time required to complete the optimal control is $T=\frac{\eta}{2q_f}\arccos{\big((1+\eta^2)/(1+\eta)^2\big)}\approx \sqrt{\eta}/2|c|$, $\eta\to +\infty$. This time-optimal method has the same leading order dependence in $\eta$ for the total time as the double-quench method but a short-pulse variation in Zeeman energy is very difficult to achieve experimentally in a spin-1 BEC system. (see Fig. \ref{fig:bang-bang}). 

For Zeeman energy values in the compact set $q \in [q_f,q_0]$ the optimal control function is piecewise constant in time, as shown in the figure. The case in which the oscillator is initially prepared in a coherent vacuum state corresponds to $q_0\to \infty$, so that the control values lie in a non compact set $q \in [q_f, \infty). $  In this case the optimal control reduces to a constant function plus a Dirac measure in time.

\bibliography{apssamp}

\end{document}